\documentclass[twocolumn,showpacs,preprintnumbers,amsmath,amssymb]{revtex4}

\usepackage{graphicx}% Include figure files
\usepackage{dcolumn}% Align table columns on decimal point
\usepackage{bm}% bold math

\begin{document}

\title{Very large magnetoresistance in lateral ferromagnetic
(Ga,Mn)As wires with nanoconstrictions}

\author{C. R\"{u}ster, T. Borzenko, C. Gould, G. Schmidt, L.W. Molenkamp}
\affiliation{Physikalisches Institut (EP3), Universit\"{a}t
W\"{u}rzburg, Am Hubland, D-97074 W\"{u}rzburg, Germany}

\author{X. Liu, T.J. Wojtowicz$^*$, J.K. Furdyna}
\affiliation{Physics Department, Notre Dame University, Notre
Dame, IN 46556}

\author{Z.G. Yu, M.E. Flatt\'{e}}
\affiliation{Dept. of Physics and Astronomy, University of Iowa,
Iowa City, IA 52242}

\date{\today}

\begin{abstract}
We have fabricated (Ga,Mn)As nanostructures in which domain walls
can be pinned by sub-10 nm constrictions. Controlled by shape
anisotropy, we can switch the regions on either side of the
constriction to either parallel or antiparallel magnetization. All
samples exhibit a positive magnetoresistance, consistent with
domain-wall trapping. For metallic samples we find a
magnetoresistance up to 8\%, which can be understood from spin
accumulation. In samples where, due to depletion at the
constriction, a tunnel barrier is formed, we observe a
magnetoresistance of up to 2000 \%.
\end{abstract}


\maketitle

Spin-valve effects involving magnetic semiconductors are of
considerable importance for applications in spintronics. Because
the carriers in these materials are holes (with strong spin-orbit
coupling), it has been difficult to observe spin dependent
scattering in the diffusive regime (GMR, or giant
magnetoresistance)\cite{akiba}. Tunnel magnetoresistance (TMR)
experiments have been more successful, yielding spin-valve effects
of several tens of percent\cite{tmr}.

It was recently pointed out\cite{flatte} that large MR effects can
be expected from domain walls (DW) in magnetic semiconductors due
to the large spin polarization in these materials. Domain wall
resistance (DWR) in ferromagnetic metals, where small effects are
typical, has been intensively investigated\cite{kentreview}. Large
MR's, however, have been observed in mechanically manipulated
nanojunctions\cite{garcia}, though magnetostriction can play a
role\cite{ansermet}.

Here we address the suggestion of Refs. \cite{flatte} in an
experimental study of DWR in the ferromagnetic semiconductor
(Ga,Mn)As. We use lateral nano-fabricated constrictions in single
domain wires to pin the DW and reduce their length\cite{bruno}.
This approach facilitates ballistic hole transport through the DW,
while the lateral fabrication technology excludes any influence of
magnetostriction effects\cite{ansermet}. We find that DWR leads to
very large spin-valve effects in both the GMR-(up to 8 \%) and
TMR-(2000 \%) regimes.

Our samples were made from a thin (19nm) epitaxial layer of
Ga$_{0.976}$Mn$_{0.024}$As, grown on a semi-insulating GaAs (001)
substrate by low temperature molecular beam epitaxy. The carrier
density from etch capacitance-voltage calibrations is about $3
\times 10^{20}$ cm$^{-3}$, and the sheet resistivity at 4.2 K is
about 4.5 k$\Omega/\square$. Assuming an effective hole mass as in
GaAs, $m^*\approx 0.5 m_o$, where $m_o$ is the free electron mass,
these values imply a Fermi energy $E_F$ of 150 meV, a Fermi
wavelength $\lambda_F$ of 6 nm, and a transport mean free path
$l_{t}$ of ca. 1 nm. The Curie temperature of the material is 65
K\cite{OhnoBook}.

\begin{figure}[h]
  \centering
  \includegraphics[width=8cm]{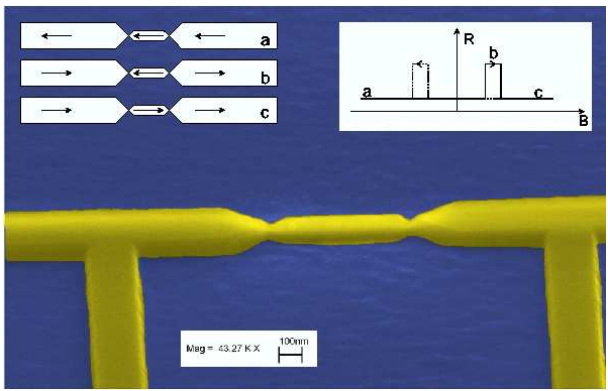}
  \caption{False-color SEM picture (side-view) of a double constriction
  showing part of the   outer wires with the voltage leads. Note
  the resist that is still present on the wire. The insets show the relative
  magnetization of the parts (left) and the resulting schematic MR
  trace for sweep-up (solid line) and sweep-down (dashed line). }
  \label{fig1}
\end{figure}

We have fabricated transport structures consisting of a central
island of 100 nm width and 500 nm length connected to two 400 nm
wide and 10 $\mu$m long wires by constrictions with widths down to
10 nm or less (Fig.\ref{fig1}). The constrictions act as pinning
centers for DW. The 400 nm wide wires are contacted by voltage-
and current-leads, allowing four-probe transport measurements. The
contact pads were defined on the (Ga,Mn)As layer by e-beam
lithography, evaporation of W and Au, and lift-off. Subsequently,
the wires and constrictions were defined by negative e-beam
lithography. Cl$_2$-based dry-etching was used to etch through the
(Ga,Mn)As, leaving (Ga,Mn)As underneath the resist and
metallization. The long axis of the island is oriented along the
[100]- (or equivalent) direction of the (Ga,Mn)As, which is near
the magnetic easy axis of this layer, as determined by SQUID and
consistent with Ref. \cite{roukes}. By leaving the resist on the
sample we can further etch it and narrow the constrictions.

We study symmetric double constriction samples in order to avoid
complications associated with thermoelectric voltages. Moreover,
the sample layout is such that the shape anisotropy causes the
(magnetically isolated) inner island to switch at different fields
than the outer wires\cite{nucl}. We verified by SQUID magnetometry
that while the wide leads switch magnetization at around 15-20 mT
(depending on lithographic parameters), the island exhibits
switching fields of order 60-90 mT. The coercive field of the
unpatterned epilayer is $\approx$ 8 mT.

MR measurements are carried out at 4.2 K in a He bath cryostat
with a superconducting magnet. The magnetic field is applied
parallel to the current direction. Four-probe DC measurements of
the magnetoresistance are performed at constant voltage using
zero-offset voltage- and current-preamplifiers. During the
measurement the magnetic field is varied from full negative
saturation of the material to full positive saturation and back.

The inset in Fig. 1 schematically describes the expected MR of our
device. Sweeping the field from large negative to positive values
causes the outer wires to switch first, inducing antiparallel
alignment of the island and the wires. In this state, DW are
present at the constrictions and the resistance is increased. At
larger positive fields the magnetization of the island is also
reversed, leaving all areas aligned in parallel and the resistance
returns to its original value. The magnetization reversal is
hysteretic, leading to the depicted spin-valve-like MR.

\begin{figure}[h]
  \centering
  \includegraphics[width=7cm]{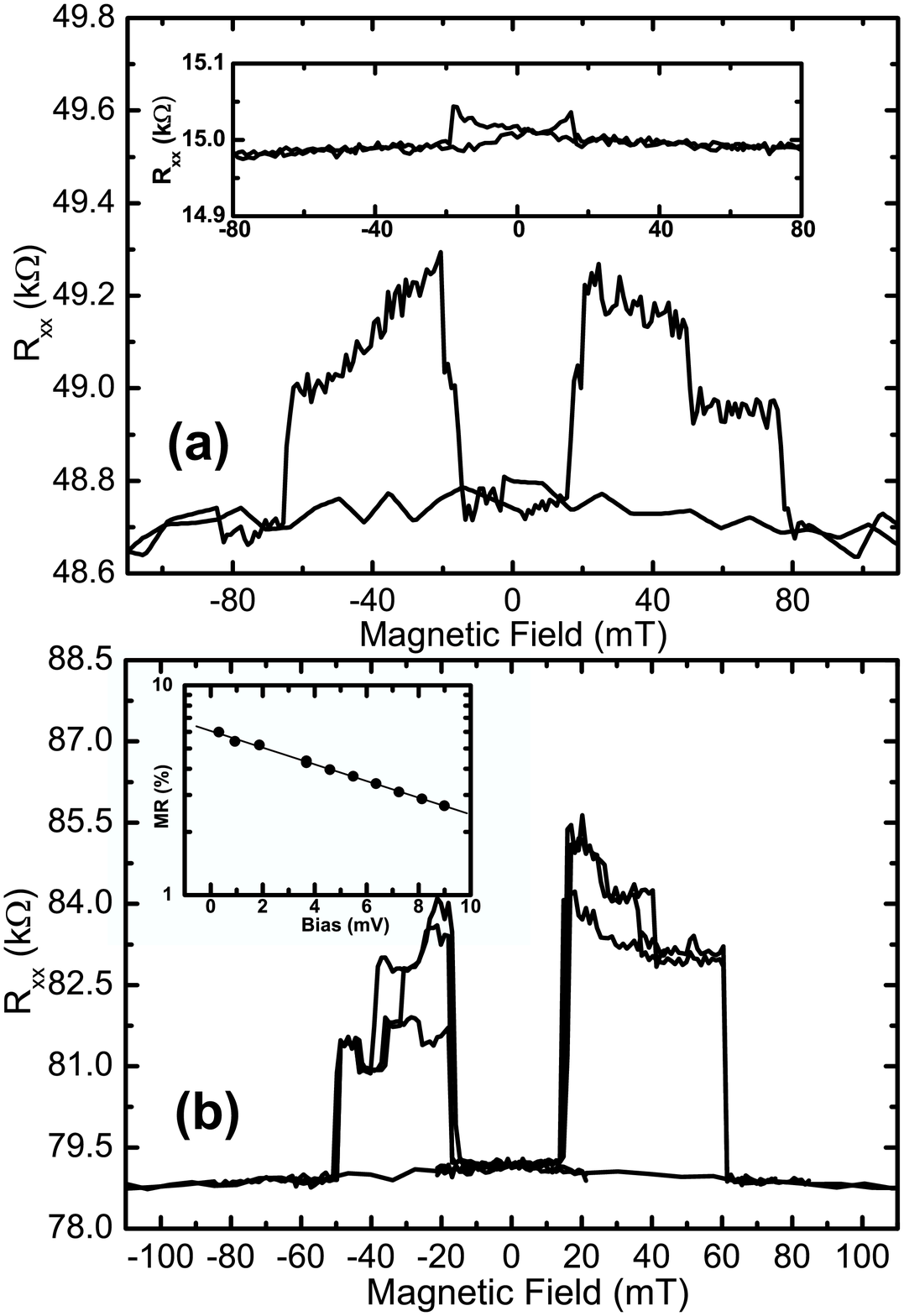}
  \caption{MR of a representative sample as fabricated (a)
  and after further etching (b). The inset in (a) shows the MR of a wire without constrictions, whereas the inset in (b) shows that the MR is reduced
  exponentially with bias voltage.}
  \label{fig2}
\end{figure}

Experimentally, all samples exhibit the expected MR. The amplitude
of the effect depends strongly on the resistance of the
constrictions. Fig. \ref{fig2}(a) shows the MR of a representative
sample with a four-terminal resistance of $\approx$48 k$\Omega$.
The MR is spin-valve-like with the correct hysteresis, and the
critical fields agree with the SQUID results. We regard this as
evidence that our design successfully incorporates both
shape-anisotropy-controlled switching and strong pinning of DW in
the constrictions. For comparison, the inset of Fig. \ref{fig2}(a)
shows the MR of a 400 nm wire without constrictions which shows
only 0.3\% bulk-like anisotropic MR, with switching at $\approx$20
mT. In Fig. \ref{fig2}(a), the maximum MR is $\approx$1.5\%.

Because the resist remains on the sample after fabrication, we can
apply a further dry etching step producing a sample with narrower
constrictions, but with approximately the same width of island and
leads. Applying this procedure to the sample of Fig.
\ref{fig2}(a), the additional etching brought the device
resistance up to $\approx$78 k$\Omega$ while the MR effect
increased to about 8\% (Fig. \ref{fig2}(b)).

The fine structure on the peaks of the as-fabricated sample was
completely reproducible, whereas for the re-etched sample it
varied from sweep to sweep as evident in the figure. This could
suggest the presence of multiple pinning sites near the
constrictions. The different pinning sites would yield different
geometrical confinement of the DW, thus altering their resistance.
Because of the extremely small dimensions realized here,
impurities and side wall roughness by etching are likely causes.

\begin{figure}[h]
  \centering
  \includegraphics[width=8cm]{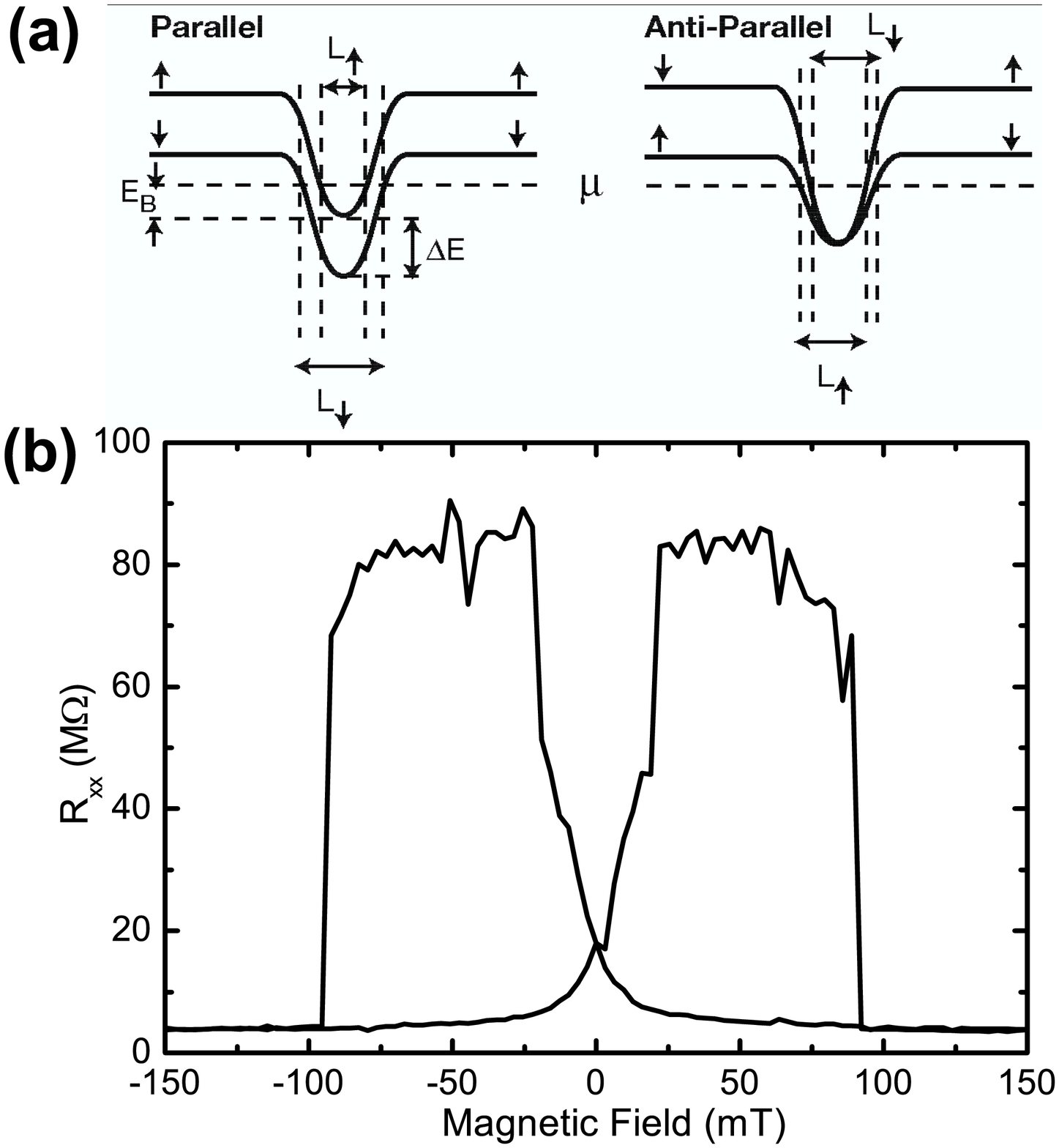}\\
  \caption{(b) MR in a sample with tunnel barriers at the
  constrictions. The top panel (a) depicts schematically the tunneling
  events taking place in the sample.}\label{fig3}
\end{figure}

A very strong increase in MR is obtained when the constrictions
are etched still further. In Fig. \ref{fig3} we plot the MR of a
sample with a zero field resistance of 4 M$\Omega$ and a positive
MR of nearly 2000\%. From the bulk resistivity of the material, we
estimate that the leads and wires in our devices contribute only
$\approx$40 k$\Omega$ to the resistance, implying that the
constructions now act as tunnel barriers. In addition, the I-V
characteristics of the sample are strongly non-linear with a
quadratic dependence of the conductance on bias that is
characteristic of tunneling transport\cite{brinkman}. This
suggests that the observed very large MR could be due to TMR. We
also note that, in contrast with the results in Fig. \ref{fig2},
we now observe a hysteretic signal around zero field (although a
major jump in resistance still occurs at the 20 mT expected from
the wide leads). This observation is also consistent with the
presence of tunnel barriers, in that these cause a magnetic
decoupling of the island and the wires. We suggest that in Fig.
\ref{fig3} the magnetization of the island no longer switches by
the introduction of a DW through a constriction but rather by
magnetic rotation, which explains the MR at zero field: If the
structure is not perfectly aligned along the easy axis, the
relatively wide wires will be magnetized slightly off-axis at zero
field while the narrow island is fully dominated by shape
anisotropy, so that the relative alignment is not fully parallel.
In wider constrictions the magnetic coupling prevents this effect
and MR is observed only at finite fields.

We can understand the above observations in a unified manner by
assuming that etching causes a gradual depletion of the carrier
density at the constrictions. Dry etching of semiconductors
damages the walls of the epilayer, and the resulting charged
impurities induce sidewall depletion of the interior of the
semiconductor, a mechanism that clearly will be most effective at
the narrowest parts of the structure, i.e. at the constrictions.
In the numerical estimates below, we assume both constrictions
have equal resistance. Note that deviations from this assumption
have only minor effects on the drawn conclusions.

As a first order approximation, we assume that the DWR is given by
the expression of Valet and Fert\cite{valet} for the
spin-accumulation-induced resistance at an abrupt junction between
two regions of opposite magnetization:
\begin{equation}
\Delta R \approx 2\beta^2 \rho^* \ell,
\end{equation}
where $\rho^*$ is the spin-symmetric bulk resistivity in the
magnetic material, $\ell$ is the length of the constriction and
$\beta$ is the spin polarization in the constriction given by
\begin{equation}
\beta = {N_{\uparrow}(E_F)v_{F\uparrow} -
N_{\downarrow}(E_F)v_{F\downarrow}\over
N_{\uparrow}(E_F)v_{F\uparrow} +
N_{\downarrow}(E_F)v_{F\downarrow}} = {E_{F\uparrow}
-E_{F\downarrow}\over E_{F\uparrow} + E_{F\downarrow}}.
\end{equation}
Here we have used that in a parabolic band model both the
densities of states at the Fermi energy
$N_{\uparrow,\downarrow}(E_F)$ and the Fermi velocities\cite{vf2}
$v_{F\uparrow,\downarrow}$  are proportional to $k_F$ (the arrows
refer to the spin subbands). From Eq. 2 one can see the evolution
of the resistance and MR with etching. We assume that etching
depletes the (Ga,Mn)As, so that $E_{F\uparrow}$ and
$E_{F\downarrow}$ are reduced, but the exchange splitting $\Delta
E = E_{F\uparrow} - E_{F\downarrow}$ remains roughly the same.
Hence the numerator of Eq. 2 does not change, but the denominator
gets smaller and the polarization increases. We now insert this
expression for the polarization into the Valet-Fert expression
(Eq. 1) for the MR. Calculating the resistance $R_c$ of the
constrictions by substracting $\approx$40 k$\Omega$ from the
device resistance, and taking a reasonable value\cite{mcd} of
$\Delta E \approx 30$ meV for the exchange splitting, we reproduce
the observed MR in Fig. 2a using the Fermi energy of 150 meV found
for the unetched sample, while the data in Fig. 2b imply a
reduction of the Fermi energy to about 90 meV. These values seem
quite reasonable, but should, given the many uncertainties and
approximations involved, only serve as a rough indication of what
may be going on in the sample. Note that the spin polarization of
some 20\% obtained from these numbers is only a lower limit
estimate of the bulk value: we know that the transport mean free
path $l_{t}$ of the holes is shorter than the dimensions of the
constriction, so we can assume that substantial spin relaxation is
taking place.

We now turn to the data in Fig. 3b, where the constrictions
clearly are in the tunneling regime. It is tempting to try to
model the observed MR in terms of Julliere's TMR
model.\cite{julliere} In order to explain a 2000 \% MR signal, the
Julliere model requires a spin polarization of the contacts of ca.
95 \%, much larger than the values found above. This large
discrepancy suggests that the model in Ref. \cite{julliere} is not
applicable here and we have therefore adopted a different approach
to modelling the tunneling regime. We also note that a model of
the MR effect due to the joint effects of both constrictions
\cite{petukhov} can be ruled out because of the long distances
between the barriers and the very short mean free path. We can
therefore safely consider a model based on two independent
barriers.

We assume that the etching process creates a shallow barrier of
parabolic shape between the two regions of (Ga,Mn)As, as shown in
Fig. 3a. We define the barrier height for the majority-spin holes
above the chemical potential $\mu$ as $E_B$. If the barrier is
very thin, such that the hole wave functions can penetrate into
the barrier region and continue to couple the Mn spins, then it is
reasonable to assume that $\Delta E$ in the barrier region is the
same as in the bulk.  This results in a barrier for minority-spin
holes that is higher ($E_B + \Delta E$) than for majority-spin
holes ($E_B$). As a consequence of the non-abrupt barrier, the
thickness of the barrier for minority-spin holes,
$L_{P\downarrow}$, will be greater than that for majority spin
holes, $L_{P\uparrow}$.

In the antiparallel situation depicted on the right, however, the
barriers for the two spin channels are the same at approximately
$E_B + \Delta E/2$, and their thicknesses are also the same,
$L_{A} \approx (L_{P\uparrow}+L_{P\downarrow})/2$. The
transmission probability $T$ through a parabolic barrier
is\cite{sze}
\begin{equation}
T = \exp\left( {-\pi m^{*(1/2)}E_H^{(1/2)}L\over 2^{3/2},
\hbar}\right)
\end{equation}
where $E_H$ and $L$ are the height and thickness of the barriers
respectively, and $\hbar$ is the reduced Planck constant. With
$m^*\approx 0.5 m_o$, we have $ T = \exp\left(-3.0
E_H^{(1/2)}L\right), $ where $E_H$ is in eV and $L$ in nm.

We now estimate the values of the parameters required to match the
experiment. Assuming that the parabolic shape of the barrier is
the same for all situations, there is a uniform relationship
between $L$ and $E_H$ of the form $L = (\alpha E_H)^{(1/2)}$,
where $\alpha$ is constant. This implies that $ T = \exp\left(-3.0
\alpha^{(1/2)} E_H\right)$. From the experiment we have $
{T_{P\uparrow}/ T_A} = 20$, and so we choose $\alpha$ to satisfy
this. With $\Delta E = 30$~meV, this yields $\alpha = 4400
$~eV$^{-2}$, independently of the barrier height for majority spin
holes $E_B$. We can estimate $E_B\approx 31$~meV from the
resistance of the constrictions. Just as a gauge, the thicknesses
of the barriers are then $11.7$~nm for the parallel majority case,
$14.3$~nm for the antiparallel case, and $16.4$~nm for the
parallel minority case. These numbers all seem reasonable.

A key element of this analysis is that the minority and majority
carriers deplete at different positions in the constriction.
Depletion at the edge of a (Ga,Mn)As film differs considerably
from depletion in the bulk, since Mn spins at the edge remain
coupled through the remaining holes to Mn spins which lie
effectively within the bulk. The presence of these nearby
bulk-like oriented Mn spins produces, through the mediating holes,
a large exchange field on the Mn spins at the edge. This in turn
induces them to order at local hole concentrations which, in the
bulk, would otherwise not lead to ferromagnetism. Hence we argue
that magnetically ordered Mn spins, producing an exchange
splitting for the holes similar to that in bulk, are present at
the edges of the sample where the local hole concentrations is
much lower than the bulk.

Finally, we turn to the observation of a voltage dependence shown
in the inset of Fig. 2b. The relative amplitude of the MR peak
decreases exponentially with increasing bias voltage V$_{bias}$. A
qualitatively similar behavior is observed for all samples with
constrictions that are not in the tunnelling regime. A bias
dependence of the current across a ferromagnetic DW was discussed
theoretically in Refs. \cite{flatte}, but not observed previously.
By analogy to a semiconducting p-n junction, an exponential
increase ($\propto \sinh(eV_{bias}/kT)$) of the (electrical)
current through the DW is expected. This should lead to an
exponential suppression of the MR signal as observed in the inset
of Fig. \ref{fig2}. However, instead of the expected slope of
e/kT, we find an activation energy of around 11 meV, or 35 kT.
Part of the discrepancy stems from the fact that not all the
applied bias drops across the constrictions (a factor of 2 in this
case), and another factor of 2 comes from the fact that our
devices have two constrictions. Nonetheless, the deviation from
Refs. \cite{flatte} is so large that we assume additional physics
is at work here. We suggest that the discrepancy may be caused by
high electric fields at the constrictions at elevated bias. In
semiconductors, such fields may cause drift effects to dominate
the transport\cite{yu}, causing a strong reduction of the
up-stream spin diffusion length. This would substantially modify
the exponential increase of the current. It would be of interest
to investigate such effects in detail, both experimentally and
theoretically.

We thank A. Brataas and C. Timm for useful discussions. This work
was supported by the BMBF, the European Commission (the FENIKS
consortium), and the DARPA SPINS program.


\begin{thebibliography}{99}
\bibitem{vf2} Eq. (2) assumes ballistic transport across the DW; in the diffusive regime one has $\beta
=(E_{F\uparrow}v_{F\uparrow}-E_{F\downarrow}v_{F\downarrow})/
(E_{F\uparrow}v_{F\uparrow} + E_{F\downarrow}v_{F\downarrow})$.
\end{thebibliography}
\end{document}